# POWER PROCESSING CIRCUITS FOR MEMS INERTIAL ENERGY SCAVENGERS

*P. D. Mitcheson, T. C. Green and E. M. Yeatman*

Department of Electrical and Electronic Engineering, Imperial College London, SW7 2AZ, U.K.

## ABSTRACT

Inertial energy scavengers are self-contained devices which generate power from ambient motion, by electrically damping the internal motion of a suspended proof mass. There are significant challenges in converting the power generated from such devices to useable form, particularly in micro-engineered variants. This paper presents approaches to this power conversion requirement, with emphasis on the cases of electromagnetic and electrostatic transduction.

## 1. INTRODUCTION

Inertial energy-scavenging devices started to be reported in the research literature about 10 years ago [1], and since then the field of inertial micro-generators has attracted much interest, and the number of active research groups in the field has grown steadily. Potential applications of such generators are in powering medical implanted devices and other types of ubiquitous computing nodes. The devices typically use a proof mass mounted on a spring suspension within a frame. When the frame is accelerated, causing relative displacement between the frame and proof mass, energy is extracted from the mechanical system by an electric damping mechanism which may be electromagnetic (typically a coil and permanent magnet) [1], electrostatic (a variable capacitor) [2] or piezoelectric (normally a cantilever bimorph structure) [3]. Most of the reported generators are based around resonant mass-spring systems, although for some applications (particularly generators designed to power medical devices) non-resonant systems can achieve higher power densities [4]. The vast majority of reported work to date has concentrated on the design and fabrication of the mass-spring system and the transducer, with many groups using MEMS technology for fabrication. Testing has normally been achieved by measuring dissipated power in a resistor. Little work has been reported on the power processing electronics, one of the functions of which is to form the interface between the transducer and the load; load circuitry requires a steady DC voltage rail and the transducer of an inertial generator does not produce a stable voltage.

The power processing electronics in a micro-generator must perform a second critical function in addition to providing a stable DC power source. There are limits on power density of an inertial energy scavenger which are primarily dependent upon the size of the generator, the motion which drives the generator frame and the architecture [4]. In order to achieve the highest possible power density under a given operating condition, it is important that the damping force is set to an optimal value, because of a trade-off between the force provided by the damper and the size of the relative motion between the mass and the frame. This value of damping is one which achieves maximum energy conversion, and thus when the transducer is operated to achieve high power densities, the electrical requirements of the damper and its characteristics are set by the need for this optimal damping force rather than simply by the electrical requirements of the load. The damping characteristics of transducer types can be altered as follows:

- Electromagnetic – the damping force can be altered by the resistance of the load connected to the coil.

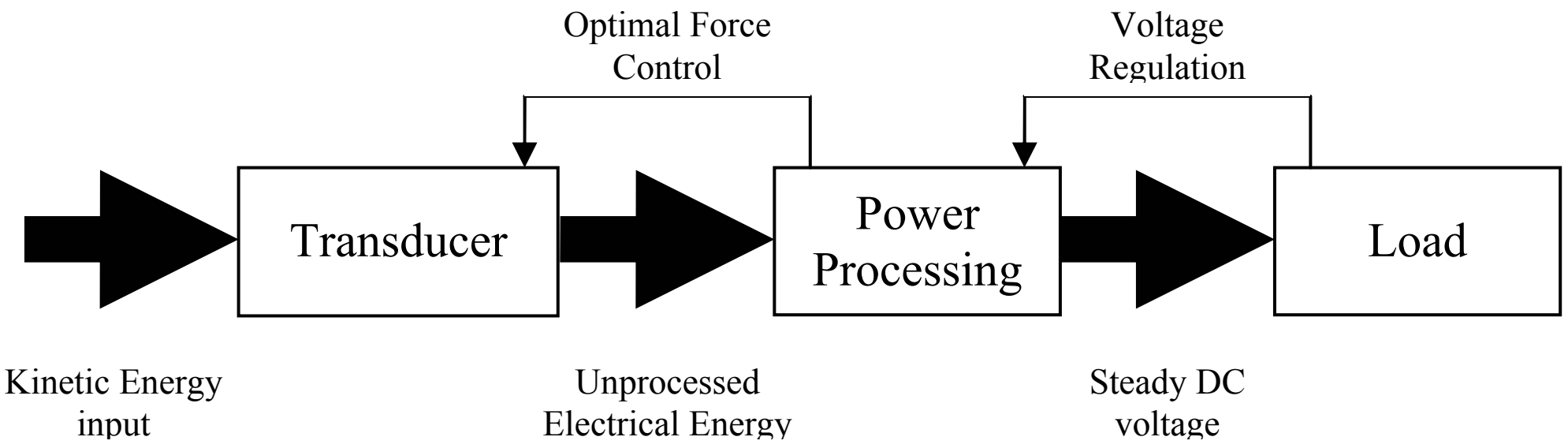

Figure 1 Block diagram of micro-generator system with power electronics

- Electrostatic – the damping force can be set by the electric field between the capacitor electrodes.
- Piezoelectric – the damping force can be altered by the resistance between the terminals of the piezoelectric cell.

The purpose of the power electronics circuitry, then, is two-fold, as shown in Figure 1 to regulate the power supply rail for the load electronics by extracting power from the transducer, and also to keep the transducer operating with the damping force that achieves the highest power density. Each of the three damper types presents different challenges in the design of the power electronics, and these add to the trade-offs in system design. Below we examine these issues for each of the three transducer types in turn.

## 2. ELECTROMAGNETIC GENERATORS

The common implementation of the electromagnetic resonant generator uses a permanent magnet and coil arrangement to provide the damping. Such a generator can also be termed a velocity damped resonant generator (VDRG), because the damping force is proportional to (and opposing) the proof mass internal velocity. This style of generator is best suited to higher frequency, low amplitude vibration sources. An illustrative case is a source vibration with an amplitude $Y_o$ = 25 μm at a frequency f = 322 Hz, used to drive a VDRG with an internal displacement limit $Z_l$ = 1 mm and proof mass m = 0.5 g. This follows the example in [5]. If we assume that the system has been tuned to operate at the resonant point, the optimal damping (for maximum power extraction) will be that which just allows the proof mass to move to its displacement limits. Thus in this case the resonant system must provide a displacement gain of 40. The optimal damping factor will be given by:

$$\zeta = \tfrac{1}{2}\frac{Y_0}{Z_l}$$

So in this case ζ = 0.05, a very lightly damped system.

The power absorbed by this optimal damper is given by [JMEMS]:

$$P_D^{Opt} = \tfrac{1}{2}v^2 D = \tfrac{1}{2}\left(\omega Z_l\right)^2\left(2m\omega_n\zeta\right)$$

where v is the proof mass velocity, D the damping coefficient, ω = 2πf, and $\omega_n$ is the resonance frequency. For the given parameter values we obtain a power of 48.7 mW.

A key design choice for the power processing is the voltage (and corresponding current) at which this power will be extracted. Conventional switch-mode circuits that include diodes must work at well above 1 V in order for the conduction power loss in the diode to be relatively small. Even with synchronous rectification with a MOSFET, it would be advantageous to operate at a relatively high voltage and low current. The counter influence is that a large active conductor length is required in the coil to achieve high voltages and the coil can become difficult to fabricate. Using a large number of turns increases the induced voltage proportionately but also increases the self inductance of the coil at something close to the square of the number of turns. A high inductance requires a long conduction period to reach the value of current corresponding to optimal damping, and this can lead to high resistive losses. Adding more conductor material to the coil (more turns of the same cross section or the same turns at greater cross section) increases the area or the length over which flux must be supported in the air gap between the magnetic materials and requires a larger volume of permanent magnet.

For a coil with an active length of $l_a$ (the length that cuts the magnetic field during vibration) and a number of turns $N$, the voltage induced in the generator is: $V_G = N B l_a\left(\omega Z_l\right)$. The maximum flux density likely to be realised in the VDRG is about 1.2 T. For a micro-engineered generator, an active length of 20 mm might be possible. This gives an induced voltage per turn of 48 mV. It is clear that if a single turn is used then very low circuit impedances will be necessary to achieve the 2A peak current required to extract 48 mW. It was considered that up to 6 turns would be feasible and voltages up to 300 mV might be achieved. This voltage is needs rectification, but is clearly too low for the use of conventional diode rectifiers. The voltage also needs to be stepped up by a ratio of about 10 for use in standard electronics.

The case examined here, although realistic, is a specific and arbitrary one, and it could be argued that increasing N is feasible, and would greatly ease the difficulties in achieving efficient conversion and regulation. However, the required N could easily be much greater in other practical cases, where the flux gradient, active length and/or operating frequency is lower, and the literature indicates that high output voltages can often not be achieved. Thus we see the low-voltage rectification and step-up requirements as being general to a large fraction of electromagnetic inertial micro-generators.

### 2.1. Proposed Dual Polarity Boost Circuit

Our proposed solution is to separately process the positive and negative half cycles of the generated voltage. Diode rectification will be replaced by alternate activation of one of two voltage boost circuits. This is a form of synchronous rectification which avoids a series connection of separate rectifier and voltage converter. To limit the step-up ratio, the two circuits will provide half the output voltage each. The target output voltage is 3.3 V and this will be provided as ±1.65V.

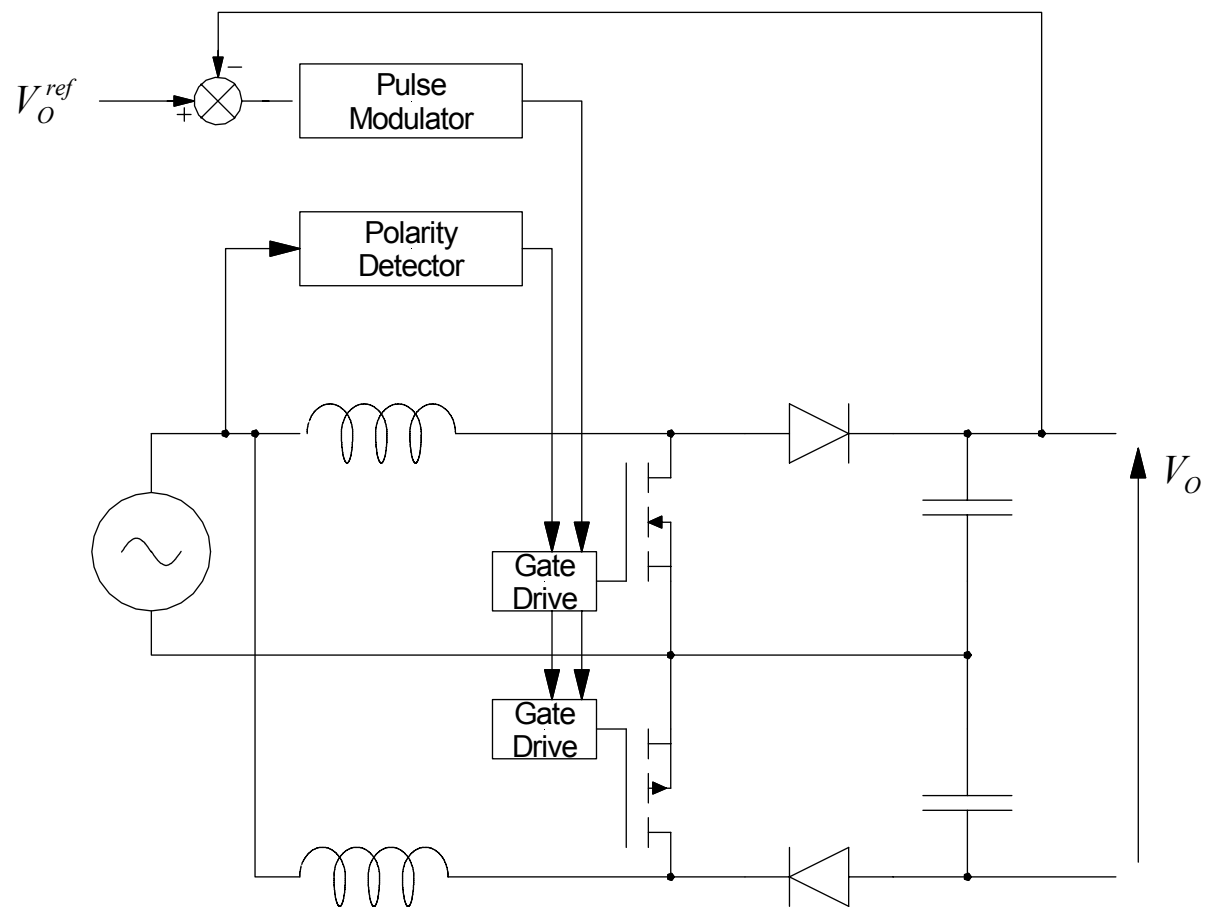

Figure 2. A dual polarity boost converter.

Figure 2 shows the two boost converter sub-circuits: one configured to produce the top half of the output voltage when the generator voltage is positive, and one configured to produce the lower half, when the generator voltage is negative. Because the generator voltage is small it is not able to forward bias the parasitic diodes of the MOSFETS. This means that in order to prevent conduction in the negative polarity boost converter when the generator voltage is positive, it is sufficient to hold off the MOSFET of the negative polarity converter. This gating of the two converters needs to be synchronised to the generator voltage. Synchronous rectification has been integrated into the boost converter so as to avoid series connection of separate rectifier and boost stages.

It is proposed to operate the boost converters in discontinuous conduction mode to avoid turn-on power loss in the MOSFET and reverse recovery effects in the diode. Several other benefits follow from this choice: relatively small passive components can be employed and a degree of resonant action can be added to the output side to manage the device parasitic capacitance. Schottky diodes have been used in this simple example but synchronously switched MOSFETs could be used instead.

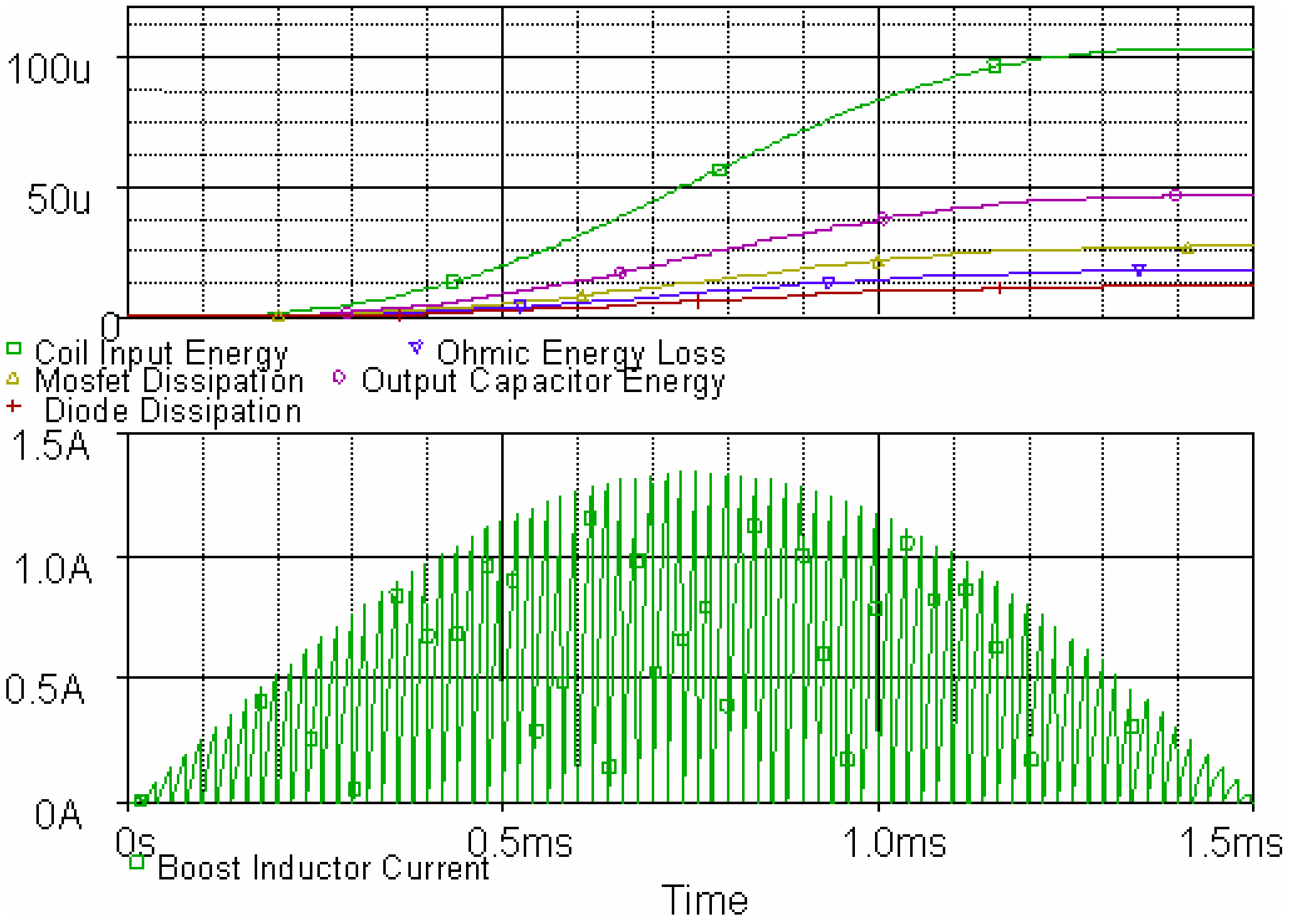

Figure 3. Spice simulation of positive half cycle. Top: accumulated energy input to the coil, output to the reservoir, and dissipated in the three main loss mechanisms. Bottom: boost inductor instantaneous current.

The generator was modelled with 4 rectangular turns of 20 mm by 4mm using 0.4 mm radius wire. The generator parameters are then: peak voltage 95 mV, self inductance 370 nH, resistance 7 mΩ, capacitance 7 pF. The main inductor was a Brooks coil of 6 turns of 0.6 mm square section wire giving: inductance 1.5 µH, resistance 28 mΩ and capacitance 31 pF. The MOSFET model was based on the commercial 2N6660 but with an area scaled by a factor of 16 (and the bonding wire resistance reduced).

Figure 3 shows the results of a spice simulation of operation over the positive half-cycle. The MOSFET was switched at 50 kHz with an on-time of 18 µs. The lower graph shows that the current drawn from the generator follows a sinusoidal envelope and the cycle-by-cycle of the current pulses reaches a peak of approximately 0.5 A. This represents 49 mW taken from the 195 mV source. The top axes show cumulative energies: input, output, resistance power losses, MOSFET power loss and diode power loss. The ratio of output energy to input energy indicates that the converter is operating at about 44% efficiency and that power loss in the MOSFET is the largest cause of inefficiency.

These results show that for low voltage electromagnetic generators with output powers in the region of 50 mW, it is possible to achieve up-conversion to useful voltages with an efficiency in the region of 50%. The circuit has been simulated with the dominant parasitic components accounted for and with reasonable device models.

## 3. ELECTROSTATIC GENERATORS

The fundamental cause of difficulty in processing the output power for constant-charge electrostatic micro generators is that they work with small amounts of charge at high voltage. The principle of operation is that a variable capacitor is charged to a relatively low voltage (the optimal value of which is dictated by the operating condition and architecture of generator [4]) at high capacitance. When the generator experiences acceleration, the capacitance of the variable capacitor falls and assuming the plates are isolated as they separate, the voltage rises. Under typical operation, the voltage generated on the plates can be of the order of a few hundred volts. This charge must be down-converted to a lower voltage in order to be suitable for powering low-power, low-voltage loads.

The energy generated is given by:

$$E = \frac{1}{2}Q^2\left[\frac{1}{C_{open}} - \frac{1}{C_{closed}}\right]$$

and thus for a given amount of charge it is necessary to achieve a high ratio between the open and closed capacitance in order to maximize energy generation. Parasitic capacitance in parallel with the generator is a major problem. Whilst parallel parasitic capacitance is likely to be negligible compared to the maximum capacitance of the generator, it will generally be non-negligible compared to the minimum (open) generator capacitance and will therefore adversely effect power generation. Therefore one of the main challenges for the circuit design task is to minimize the parasitic capacitance connected to the generator.

The circuit of Fig. 4 shows a buck converter circuit which has previously been simulated using the Silvaco finite element device simulator [6] using custom designed semiconductor devices rated for high voltage blocking (around 250 V), low off-state leakage and low junction capacitances. This circuit was initially investigated because it appears to be the simplest method of down-converting the high voltage on the generator. The simulations in Silvaco allowed the conversion efficiency of the converter to be evaluated using a mixed-mode finite element/lumped element simulation for the devices and passive components respectively.

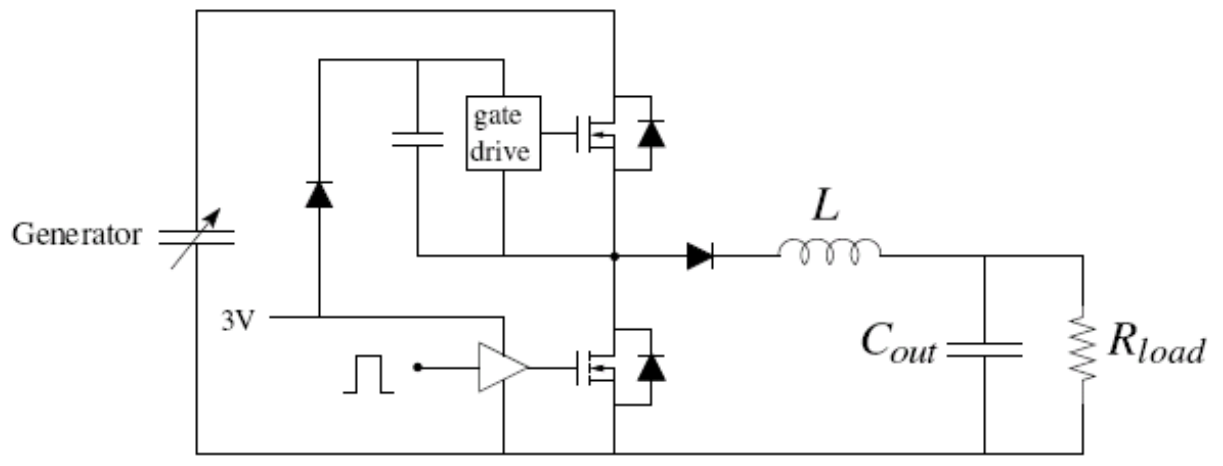

Figure 4. Modified buck converter.

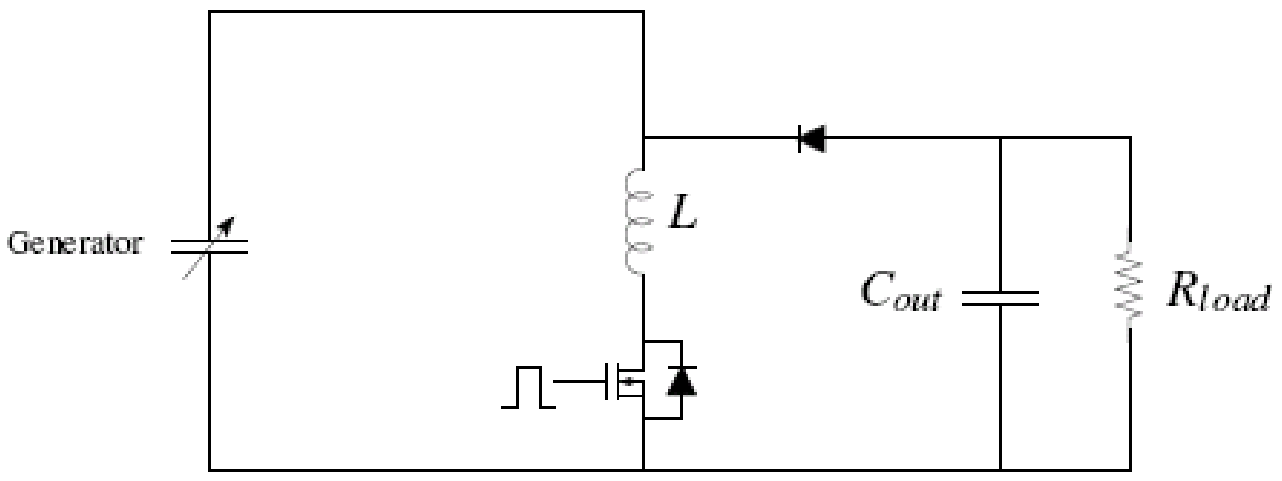

Figure 5. Modified flyback converter.

The overall effectiveness of a micro-generator is more complex than just the efficiency of the power processing circuitry, and has been defined as a product of several terms [6]. Two of the most important are the generation efficiency $n_{gen}$ and the conversion efficiency $n_{conv}$. These terms are defined as follows:

$$n_{gen} = \frac{E_{open}}{W_{field} + E_{closed}}$$

$$n_{conv} = \frac{E_{out}}{E_{open}}$$

where $E_{open}$ is the energy stored on the moving plate capacitor at minimum capacitance, $E_{closed}$ the priming energy on the capacitor at maximum capacitance, $W_{field}$ is the amount of work that could have been done against the electric field as the plates separate and $E_{out}$ is the energy available after processing by the converter.

In the buck converter circuit the depletion layer capacitance of the blocking junction of the high-side MOSFET forms a parasitic capacitance in parallel with the generator capacitor as the generator voltage rises. Energy stored in this depletion layer capacitance is lost when the MOSFET is turned on.

Parasitic capacitance in parallel with the generation capacitor will reduce $n_{gen}$, and switching and conduction losses in the converter will reduce $n_{conv}$. Increasing the cross sectional area of the MOSFET and diode will tend to increase $n_{conv}$ but decrease $n_{gen}$, because of the associated additional parasitic capacitance.

Models of the custom designed semiconductor devices were created in PSpice so that $n_{conv}$ and $n_{gen}$ could be simultaneously evaluated. The results are shown in Fig. 6 where the generator efficiency is shown as the product of $n_{gen}$ and $n_{conv}$. The number of cells of refers to the number of 0.015 mm$^2$ cells that were used for the MOSFET and the diode. As can be seen, increasing the number of cells increases the conversion efficiency but reduces the generation efficiency. From this preliminary study, ten cells appears to give the highest overall generation efficiency.

An additional energy loss mechanism associated with the buck converter circuit is a shoot-through current which reverse biases the blocking junction in the low side MOSFET. The high-side gate drive is also non-trivial to design. A possible solution to these two problems is to use a modified version of the flyback converter, shown in Fig. 5. An isolated flyback converter has been suggested in [7], although details of the device parasitics are not presented. However, when this circuit was evaluated in PSpice, although the conversion efficiency of the flyback converter can be higher at higher cross sectional areas of device (because of the lack of shoot-through current), the additional parasitic capacitance from the diode reduces the generation efficiency too quickly, and overall, the buck converter achieves a higher efficiency.

## 4. PIEZOELECTRIC GENERATORS

Piezoelectric devices are attractive from a power processing point of view, as they can produce voltages in some practical micro-generator applications [3] which can be processed with off the shelf semiconductor devices. However, obtaining a high damping force can be difficult with piezoelectric devices operating at low frequency due to internal leakage, and to limitations on the practical geometry and dimensions, most such devices using bimorph cantilevers. Unlike the electromagnetic and electrostatic cases, for piezoelectric generators a number of authors have described power processing circuits for converting the output to useful form.

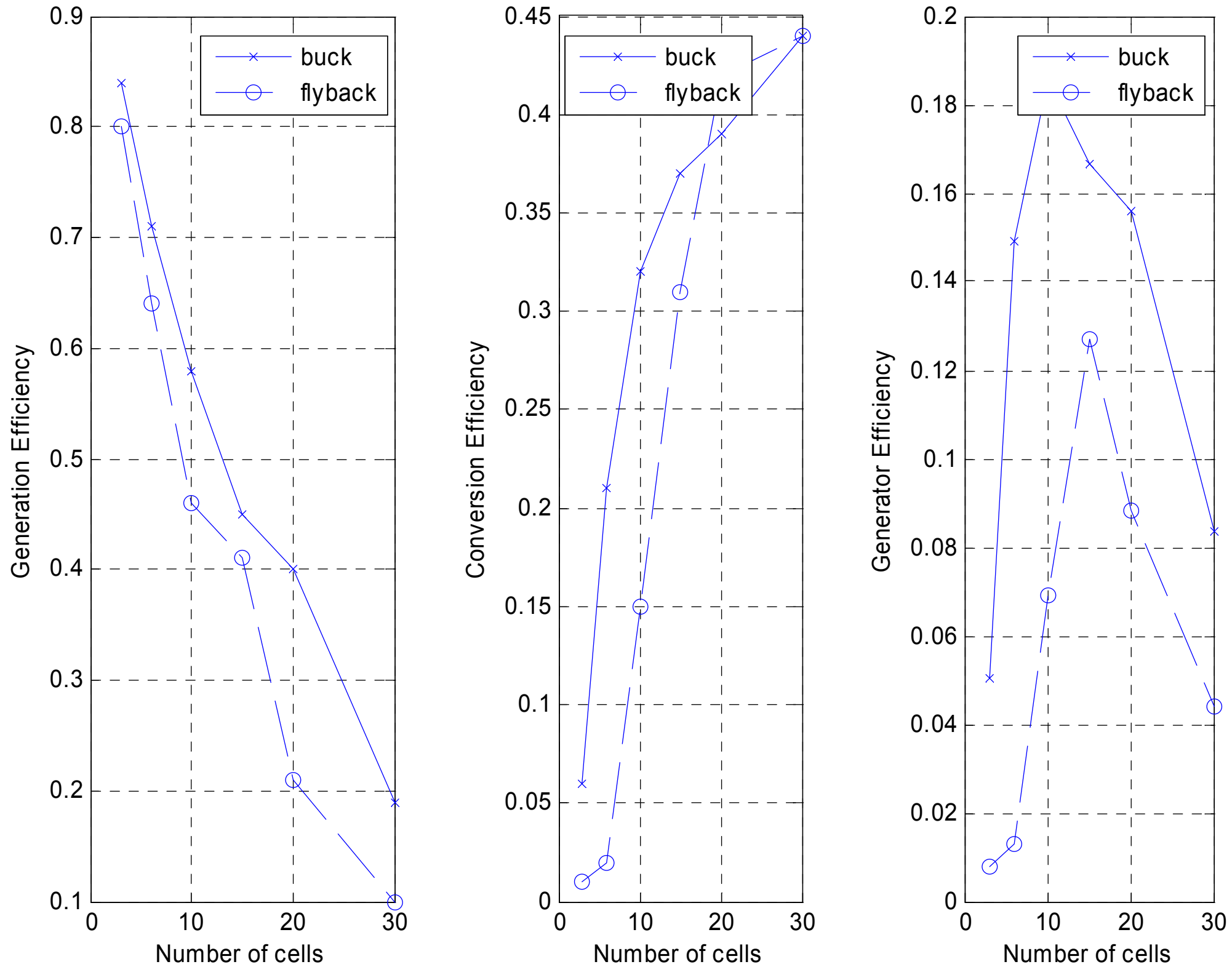

Figure 6. Effectiveness and efficiencies of an electrostatic micro-generator.

For example, Ottman *et al.* designed an optimised power processing circuit for a piezoelectric transducer [8]. In this case relatively large voltages were obtained from the transducer (up to 100 V), so that full-wave diode rectification was practical, followed by a conventional DC-DC step-down convertor. The use of duty cycle to vary the damping factor on the transducer was demonstrated.

Roundy *et al.* [3] and Ottman *et al.*[8] have both shown that piezoelectric generators can achieve higher power densities when driving resistive loads than when they are connected to a simple power supply consisting of a bridge rectifier and smoothing capacitor. We have shown that the same is true for electromagnetic devices [4], and the circuit presented in section 2 satisfies this requirement.

## 5. CONCLUSIONS

Electromagnetic and constant-charge electrostatic inertial generators each present significant challenges for output conversion and regulation, because of low and high output voltages respectively, and the need to achieve high efficiency in both cases despite high sensitivity to parasitics. We have presented circuit topologies for both these cases, and simulated them with realistic device models. In both cases acceptable efficiencies could be obtained. Since the proposed circuits each allow the effective load on the generator to be varied using switching duty cycles, they also present a convenient mechanism for dynamically optimising the load to extract maximum power under varying source motion.